\input{aipcheck}

\documentclass[
    ,final            
  ]
  {aipproc}
  
\usepackage{scalefnt}

\layoutstyle{8x11double}

\begin{document}

\title{Masses of Axial-Vector Resonances in a Linear Sigma Model with \boldmath $N_f=3$}


\author{Denis Parganlija}{
  address={Institute for Theoretical Physics, Johann Wolfgang Goethe University, Max-von-Laue-Str. 1, D--60438 Frankfurt am Main, Germany},
  email={parganlija@th.physik.uni-frankfurt.de}
}

\iftrue
\author{Francesco Giacosa}{
  address={Institute for Theoretical Physics, Johann Wolfgang Goethe University, Max-von-Laue-Str. 1, D--60438 Frankfurt am Main, Germany},
}


\author{P\'{e}ter Kov\'{a}cs}{
  address={Research Institute for Particle and Nuclear Physics, H--1525 Budapest, POB 49, Hungary},
}

\iftrue
\author{Gy\"{o}rgy Wolf}{
  address={Research Institute for Particle and Nuclear Physics, H--1525 Budapest, POB 49, Hungary},
}

\fi

\begin{abstract}
We discuss an $N_f=3$ linear sigma model with vector
and axial-vector mesons (extended Linear Sigma Model - eLSM).
We present
first results regarding the masses of axial-vector mesons determined from the
extended model.
\end{abstract}

\maketitle


\section{Introduction}

The vacuum phenomenology of low-energy mesons can be described in a variety
of approaches using the linear \cite{gellmannlevy} and non-linear \cite{weinberg} realisations of the chiral symmetry. The linear realisation of
the chiral symmetry (linear sigma model) has, for example, been used in
Refs.\ \cite{Paper1,Susanna} in order to describe non-strange hadrons in the energy
region up to approximately 1.7 GeV (see also Ref.\ \cite{Lenaghan} and Refs.\ therein).
However, this energy region contains further experimentally well-established
resonances \cite{PDG}, such as those composed solely of strange quarks, or
those with an admixture of strange quark fields. These resonances are
important for the description of meson phenomenology in vacuum and at
non-zero temperatures and are expected to play an important role in the
restoration of the chiral symmetry. Therefore, a more complete
phenomenological description of mesons requires an extension of the model to 
$N_{f}=3$. Additionally, the data ragarding strange mesons is both
precise and abundant thus offering more constraints for a phenomenological
approach than the data regarding the non-strange mesons. In Ref.\ \cite{Krakau}, we have outlined such an extension of the $N_{f}=2$ linear sigma
model with vector and axial-vector degrees of freedom, presented in Ref.\ 
\cite{Paper1}, to $N_{f}=3$ (extended Linear Sigma Model - eLSM). In this paper, we report on first results
regarding axial-vector meson masses from eLSM.\newline
The paper is organised as follows: in Sec.\ 2 we present the model
Lagrangian and its implications and in Sec.\ 3 we summarise our results.

\section{The model}

We use an $N_{f}=3$ linear sigma model with global chiral invariance in the
following form \cite{Paper1, Krakau}: 
\begin{eqnarray}
\mathcal{L} &=&\mathrm{Tr}[(D^{\mu }\Phi )^{\dagger }(D^{\mu }\Phi
)]-m_{0}^{2}\mathrm{Tr}(\Phi ^{\dagger }\Phi )  \nonumber  \label{Lagrangian}
\\
&&-\lambda _{1}[\mathrm{Tr}(\Phi ^{\dagger }\Phi )]^{2}-\lambda _{2}\mathrm{%
Tr}(\Phi ^{\dagger }\Phi )^{2}  \nonumber \\
&&-\frac{1}{4}\mathrm{Tr}[(L^{\mu \nu })^{2}+(R^{\mu \nu })^{2}]+\frac{%
m_{1}^{2}}{2}\mathrm{Tr}[(L^{\mu })^{2}+(R^{\mu })^{2}]  \nonumber \\
&&+\mathrm{Tr}[H(\Phi +\Phi ^{\dagger })]+c(\det \Phi +\det \Phi ^{\dagger })
\nonumber \\
&&-2ig_{2}(\mathrm{Tr}\{L_{\mu \nu }[L^{\mu },L^{\nu }]\}+\mathrm{Tr}%
\{R_{\mu \nu }[R^{\mu },R^{\nu }]\})  \nonumber \\
&&-2g_{3}\left[ \mathrm{Tr}\left( \left\{ \partial _{\mu }L_{\nu }+\partial
_{\nu }L_{\mu }\right\} \{L^{\mu },L^{\nu }\}\right) \right.  \nonumber \\
&&\left. +\mathrm{Tr}\left( \left\{ \partial _{\mu }R_{\nu }+\partial _{\nu
}R_{\mu }\right\} \{R^{\mu },R^{\nu }\}\right) \right]  \nonumber \\
&&+\frac{h_{1}}{2}\mathrm{Tr}(\Phi ^{\dagger }\Phi )\mathrm{Tr}[(L^{\mu
})^{2}+(R^{\mu })^{2}]  \nonumber \\
&&+h_{2}\mathrm{Tr}[(\Phi R^{\mu })^{2}+(L^{\mu }\Phi )^{2}]+2h_{3}\mathrm{Tr%
}(\Phi R_{\mu }\Phi ^{\dagger }L^{\mu }),  \nonumber \\
&&
\end{eqnarray}%
where 
\begin{equation}
\scalefont{0.81}\Phi =\frac{1}{\sqrt{2}}\left( 
\begin{array}{ccc}
\frac{(\sigma _{N}+a_{0}^{0})+i(\eta _{N}+\pi ^{0})}{\sqrt{2}} & 
a_{0}^{+}+i\pi ^{+} & K_{S}^{+}+iK^{+} \\ 
a_{0}^{-}+i\pi ^{-} & \frac{(\sigma _{N}-a_{0}^{0})+i(\eta _{N}-\pi ^{0})}{%
\sqrt{2}} & K_{S}^{0}+iK^{0} \\ 
K_{S}^{-}+iK^{-} & {\bar{K}_{S}^{0}}+i{\bar{K}^{0}} & \sigma _{S}+i\eta _{S}%
\end{array}%
\right)   \normalsize \label{Phi}
\end{equation}%
%
%
%
is a matrix containing the scalar and pseudoscalar degrees of freedom and 
\begin{eqnarray}
\scalefont{0.67} L^{\mu } =\frac{1}{\sqrt{2}}\left( 
\begin{array}{ccc}
\frac{\omega _{N}+\rho ^{0}}{\sqrt{2}}+\frac{f_{1N}+a_{1}^{0}}{\sqrt{2}} & 
\rho ^{+}+a_{1}^{+} & K^{\star +}+K_{1}^{+} \\ 
\rho ^{-}+a_{1}^{-} & \frac{\omega _{N}-\rho ^{0}}{\sqrt{2}}+\frac{%
f_{1N}-a_{1}^{0}}{\sqrt{2}} & K^{\star 0}+K_{1}^{0} \\ 
K^{\star -}+K_{1}^{-} & {\bar{K}}^{\star 0}+{\bar{K}}_{1}^{0} & \omega
_{S}+f_{1S}%
\end{array}%
\right) ^{\mu }{\normalsize ,}  \nonumber \\
\scalefont{0.67} R^{\mu } =\frac{1}{\sqrt{2}}\left( 
\begin{array}{ccc}
\frac{\omega _{N}+\rho ^{0}}{\sqrt{2}}-\frac{f_{1N}+a_{1}^{0}}{\sqrt{2}} & 
\rho ^{+}-a_{1}^{+} & K^{\star +}-K_{1}^{+} \\ 
\rho ^{-}-a_{1}^{-} & \frac{\omega _{N}-\rho ^{0}}{\sqrt{2}}-\frac{%
f_{1N}-a_{1}^{0}}{\sqrt{2}} & K^{\star 0}-K_{1}^{0} \\ 
K^{\star -}-K_{1}^{-} & {\bar{K}}^{\star 0}-{\bar{K}}_{1}^{0} & \omega
_{S}-f_{1S}%
\end{array}%
\right) ^{ \scalefont{0.67} \mu }  \label{LR}
\end{eqnarray}%
are, respectively, the left-handed and right-handed matrices containing the
vector and axial-vector degrees of freedom. Also, $D^{\mu }\Phi =\partial
^{\mu }\Phi -ig_{1}(L^{\mu }\Phi -\Phi R^{\mu })$ is the covariant
derivative; $L^{\mu \nu }=\partial ^{\mu }L^{\nu }-\partial ^{\nu }L^{\mu }$
and $R^{\mu \nu }=\partial ^{\mu }R^{\nu }-\partial ^{\nu }R^{\mu }$ are,
respectively, the left-handed and right-handed field strength tensors; the
term Tr$[H(\Phi +\Phi ^{\dagger })]$ [$H=1/2\,\mathrm{diag}(h_{0N},h_{0N},%
\sqrt{2}h_{0S})$, $h_{0N}=const.$, $h_{0S}=const.$] explicitly breaks chiral
symmetry due to nonzero quark masses, and the term $c\,(\det \Phi +\det \Phi
^{\dagger })$ describes the $U(1)_{A}$ anomaly \cite{Hooft}.

As in Ref.\ \cite{Paper1}, in the non-strange sector, we assign the fields $%
\vec{\pi}$ and $\eta _{N}$ to the pion and the $SU(2)$ counterpart of the $%
\eta $ meson, $\eta _{N}\equiv (\overline{u}u+\overline{d}d)/\sqrt{2}$. The
fields $\omega _{N}^{\mu }$ and $\vec{\rho}^{\mu }$ represent the $\omega
(782)$ and $\rho (770)$ vector mesons, respectively, and the fields $%
f_{1N}^{\mu }$ and $\vec{a}_{1}^{\mu }$ represent the $f_{1}(1285)$ and $%
a_{1}(1260)$ mesons, respectively. In the strange sector, we assign the $K$
fields to the kaons; the $\eta _{S}$ field is the strange contribution to
the physical $\eta $ and $\eta ^{\prime }$ fields; the $\omega _{S}$, $%
f_{1S} $, $K^{\star }$ and $K_{1}$ fields correspond to the $\phi (1020)$, $%
f_{1}(1420)$, $K^{\star }(892)$, and $K_{1}(1270)$ mesons, respectively. In
accordance with Ref.\ \cite{Paper1}, where the scalar ${\bar{q}}q$ states
were found in the energy region above 1 GeV, we assign the scalar kaon $%
K_{S} $ to the physical $K_{0}^{\star }(1430)$ state. The preliminary
results from our extended model, Eq.\ (\ref{Lagrangian}), seem to point to
the predominantly strange and non-strange sigma states to be above 1 GeV as
well \cite{Krakau} (these states arise from the mixing of the pure
quarkonium state $\sigma _{N}$ and an the pure glueball state $\sigma _{S}$).%
\newline
In order to implement spontaneous \cite{Paper1,Goldstone1}
breaking of the chiral symmetry in vacuum by the quark condensate, we shift
the $\sigma _{N}$ and $\sigma _{S}$ fields by their respective vacuum
expectation values $\phi _{N}$ and $\phi _{S}$.\newline
The spontaneous symmetry breaking results in $\eta _{N}$-$f_{1N}$ and $\vec{%
\pi} $-$\vec{a}_{1}$ mixings: $-g_{1}\phi _{N}(f_{1N}^{\mu }\partial _{\mu
}\eta _{N}+{\vec{a}_{1}^{\mu }}\cdot \partial _{\mu }{\vec{\pi}})$ \cite%
{Paper1} as well as in $\eta _{S}$-$f_{1S}$, $K_{S}$-$K^{\star }$, and $K$-$%
K_{1}$ mixings: $-\sqrt{2}g_{1}\phi _{S}f_{1S}^{\mu }\partial _{\mu }\eta
_{S}$, $ig_{1} ( \sqrt{2}\phi _{S}-\phi _{N}) ( {\bar{K}}^{\star \mu
0}\partial _{\mu }{K_{S}^{0}+}K^{\star \mu -}\partial _{\mu }{K}_{S}^{+}) /2
+ ig_{1}( \phi _{N}-\sqrt{2}\phi _{S}) ( K^{\star \mu 0}\partial _{\mu }{%
\bar{K}_{S}^{0}+}K^{\star \mu +}\partial _{\mu }{K}_{S}^{-}) /2$ and $%
-g_{1}( \phi _{N}+\sqrt{2}\phi _{S}) ( K_{1}^{\mu 0}\partial _{\mu }{\bar{K}%
^{0}+}K_{1}^{\mu +}\partial _{\mu }K^{-}/2 ) + \mathrm{h.c.} $,
respectively. Note that our Lagrangian is real despite the imaginary $K_{S}$-%
$K^{\star }$ coupling because the $K_{S}$-$K^{\star }$ mixing term is equal
to its hermitian conjugate and therefore real.

The mixing terms are removed similarly to the way described in Ref.\ \cite%
{Paper1}, with suitable shifts of the vector field $K^{\star }$ and the
axial-vector fields concerned: $f_{1N,S}^{\mu }\rightarrow f_{1N,S}^{\mu
}+w_{f_{1N,S}}\partial ^{\mu }\eta _{N,S}$; ${\vec{a}_{1}^{\mu }}\rightarrow 
{\vec{a}_{1}^{\mu }+}w_{a_{1}}\partial ^{\mu }{\vec{\pi}}$; $K^{\star \mu
0}\rightarrow K^{\star \mu 0}+w_{K^{\star }}\partial ^{\mu }{K_{S}^{0}}$; $%
K^{\star \mu +}\rightarrow K^{\star \mu +}+w_{K^{\star }}\partial ^{\mu }{%
K_{S}^{+}}$; ${\bar{K}}^{\star \mu 0}\rightarrow {\bar{K}}^{\star \mu
0}+w_{K_{\star }}^{\ast }\partial ^{\mu }{\bar{K}_{S}^{0}}$; $K^{\star \mu
-}\rightarrow K^{\star \mu -}+w_{K_{\star }}^{\ast }\partial ^{\mu }{%
K_{S}^{-}}$; $K_{1}^{\mu 0}\rightarrow K_{1}^{\mu 0}+w_{K_{1}}\partial ^{\mu
}{K^{0}}$ (and h.c. for $K_{1}$), where the $w$ constants are defined in
such a way that the mentioned mixing terms are removed from the Lagrangian: $%
w_{f_{1N}}=w_{a_{1}}=g_{1}\phi _{N}/m_{a_{1}}^{2}$ (one obtains $%
w_{f_{1N}}=w_{a_{1}}$, as in Ref.\ \cite{Paper1}), $w_{f_{1S}}=$ $\sqrt{2}%
g_{1}\phi _{S}/m_{f_{1S}}^{2}$, $w_{K^{\star }}=ig_{1}(\phi _{N}-\sqrt{2}%
\phi _{S})/(2m_{K^{\star }}^{2})$ and $w_{K_{1}}=$ $g_{1}(\phi _{N}+\sqrt{2}%
\phi _{S})/(2m_{K_{1}}^{2})$.

Subsequently, as in Ref.\ \cite{Paper1}, the fields $\eta _{N,S}$, $\vec{\pi}
$, $K_{S}$ and $K$ are no longer canonically normalised. In order to obtain
canonical normalisation, we introduce renormalisation constants
(coefficients) of these wave functions labelled $Z_{\eta _{N,S}}$ for $\eta
_{N,S}$, $Z_{\pi }$ for ${\vec{\pi}}$, $Z_{K_{S}}$ for $K_{S}$\ and $Z_{K}$
for $K$ (note that these coefficients do not contain loop corrections and can thus have a value larger than one, see Table \ref{Table1} for the values of $Z_\pi$). We obtain the following formulas:

\begin{eqnarray}
Z_{\pi } &\equiv &Z_{\eta _{N}}=\frac{m_{a_{1}}}{\sqrt{%
m_{a_{1}}^{2}-g_{1}^{2}\phi_{N}^{2}}} \\
Z_{\eta _{S}} &=&\frac{m_{f_{1S}}}{\sqrt{m_{f_{1S}}^{2}-2g_{1}^{2}\phi
_{S}^{2}}} \\
Z_{K} &=&\frac{2m_{K_{1}}}{\sqrt{4m_{K_{1}}^{2}-g_{1}^{2}(\phi _{N}+\sqrt{2}%
\phi _{S})^{2}}}  \label{ZK1} \\
Z_{K_{S}} &=&\frac{2m_{K_{\star }}}{\sqrt{4m_{K_{\star }}^{2}-g_{1}^{2}(\phi
_{N}-\sqrt{2}\phi _{S})^{2}}}.
\end{eqnarray}%
For the non-strange and strange condensates, we then have $\phi _{N}=Z_{\pi
}f_{\pi }$ \cite{Paper1} and analogously $\phi _{S}=Z_{K}f_{K}/\sqrt{2}$,
where $f_{\pi }=92.4$ MeV and $f_{K}=155$ MeV $/\sqrt{2}$ are, respectively,
the pion and kaon decay constants.

Additionally to Eq.\ (\ref{ZK1}), we obtain two more formulas for $Z_{K}$ from $m_{f_{1S}}^{2}-m_{\omega _{S}}^{2}$ and $m_{K_{1}}^{2}-m_{K^{%
\star }}^{2}$: 
\begin{eqnarray}
Z_{K} &=&\frac{1}{f_{K}}\sqrt{\frac{m_{f_{1S}}^{2}-m_{\omega _{S}}^{2}}{%
g_{1}^{2}(Z_{\pi })-h_{3}(Z_{\pi })}}  \label{ZK2} \\
Z_{K} &=&\frac{m_{K_{1}}^{2}-m_{K^{\star }}^{2}}{Z_{\pi }f_{\pi
}f_{K}[g_{1}^{2}(Z_{\pi })-h_{3}(Z_{\pi })]}.  \label{ZK3}
\end{eqnarray}%
Therefore, in order to be consistent, the values of $Z_{K}$ have to
simultaneously fulfill three equations: (\ref{ZK1}), which is the definition
of $Z_{K}$, (\ref{ZK2}) and (\ref{ZK3}). We note that Eqs.\ (\ref{ZK1}), (%
\ref{ZK2}) and (\ref{ZK3}), in addition to $m_{\omega _{S}}$, $m_{f_{1S}}$, $%
m_{K^{\star }}$ and $m_{K_{1}}$, also contain $m_{\rho }$ and $m_{a_{1}}$
present in parameters $g_{1}$ and $h_{3}$ (see Ref.\ \cite{Paper1}): 
\begin{eqnarray}
g_{1}(Z_{\pi }) &=&\frac{m_{a_{1}}}{Z_{\pi }f_{\pi }}\sqrt{1-\frac{1}{Z_{\pi
}^{2}}}  \label{g1} \\
h_{3}(Z_{\pi }) &=&\frac{m_{a_{1}}^{2}}{Z_{\pi }^{2}f_{\pi }^{2}}\left( 
\frac{m_{\rho }^{2}}{m_{a_{1}}^{2}}-\frac{1}{Z_{\pi }^{2}}\right) .
\label{h3}
\end{eqnarray}

Hence, we need to determine masses that are to be implemented in the
Eqs.\ (\ref{ZK1}), (\ref{ZK2}) and\ (\ref{ZK3}), i.e., the masses that
should correspond to the experimental data (up to loop corrections to our tree-level masses,
with corrections not expected to be large in the case of our resonances). The $\rho $ and $%
K^{\star }$ states are well-established quarkonia \cite{Pelaez}; the states
currently present in our model are ${\bar{q}}q$ states \cite{Paper1} and
thus we set the $\rho $ and $K^{\star }$ masses to the PDG values: $m_{\rho
}=775.49$ MeV and $m_{K^{\star }}=891.66$ MeV. We assign our $\omega
_{S}\equiv \bar{s}s$ state to the physical $\varphi (1020)$ resonance
because this resonance is known to be predominantly an $\bar{s}s$ field,
although with a small admixture of the non-strange quarks. Our Lagrangian
does not implement $\bar{s}s$ - $\bar{n}n$ mixing in the isosinglet vector
channel and thus, as a first approximation, we set the $\omega _{S}$ mass to
the PDG value: $m_{\omega _{S}}=1019.455$ MeV.

Given the assignment of the states in our model to the physical states, one
would usually also set $m_{f_{1S}}=1426.4 \pm 0.9$ MeV, $m_{K_{1}}=1272 \pm 7$ MeV and $%
m_{a_{1}}=1230 \pm 40$ MeV \cite{PDG}. The latter value is merely an "educated guess" \cite{PDG}.
Therefore, in order to simultaneously fulfill Eqs.\ (\ref{ZK1}%
), (\ref{ZK2}) and\ (\ref{ZK3}), we can relax the interval for $m_{a_{1}}$
and search for a suitable value in the region 1.1 - 1.3 GeV while retaining
the values of $m_{K_{1}}$ and $m_{f_{1S}}$ in the vicinity of the PDG data.
In this way, the hypothesis of $K_{1}(1270)$ and $f_{1}(1420)$ as
predominantly $\bar{q}q$ states is tested; obtaining their masses in the
vicinity of the experimental data would be an indication that this
hypothesis is justified. Note also that it is actually not possible to
simultaneously fulfill Eqs.\ (\ref{ZK1}), (\ref{ZK2}) and\ (\ref{ZK3}) if $%
m_{f_{1S}}$, $m_{K_{1}}$ and $m_{a_{1}}$ are set to their exact respective
values quoted by the PDG.

We broaden the interval $Z_\pi = 1.67 \pm 0.2$ found in Ref.\ \cite{Paper1} to
$1.1\leq Z_{\pi }\leq 1.9$ in order to obtain the axial-vector masses for more general parameter values.
Subsequently, by enforcing the equality of the three Eqs.\ (\ref{ZK1}), (\ref{ZK2})
and (\ref{ZK3}), obtain constraints on $m_{K_{1}}$, $m_{f_{1S}}$ and $%
m_{a_{1}}$. We vary $m_{a_{1}}$ between 1.1 and 1.3 GeV and look for $%
m_{K_{1}}$ and $m_{f_{1S}}$ that are as close as possible to the PDG data [$%
m_{K_{1}}^{\mathrm{PDG}}=(1272\pm 7)$ MeV and $m_{f_{1S}}^{\mathrm{PDG}%
}=(1426.4\pm 0.9)$ MeV]. We find the results presented in Table \ref{Table1}.

\begin{table}[h] \centering%
\begin{tabular}{|c|c|c|c|}
\hline
$Z_{\pi }$ & $m_{a_{1}}$ (MeV) & $m_{K_{1}}$ (MeV) & $m_{f_{1S}}$ (MeV) \\ 
\hline
1.1 & 1142 & 1276 & 1423 \\ \hline
1.2 & 1144 & 1276 & 1421 \\ \hline
1.3 & 1147 & 1277 & 1420 \\ \hline
1.4 & 1152 & 1279 & 1419 \\ \hline
1.5 & 1157 & 1281 & 1418 \\ \hline
1.6 & 1163 & 1283 & 1416 \\ \hline
1.7 & 1170 & 1285 & 1413 \\ \hline
1.8 & 1180 & 1289 & 1411 \\ \hline
1.9 & 1191 & 1292 & 1406 \\ \hline
\end{tabular}
\caption{Values of $m_{a_1}$, $m_{f_{1S}}$ and $m_{K_1}$ that, for a given
$Z_\pi$, lead to the same value of $Z_K$ from Eqs.\ (\ref{ZK1}),
(\ref{ZK2}) and\ (\ref{ZK3}).}\label{Table1}%
\end{table}%

Thus, $m_{a_{1}}$ is about 50-100 MeV smaller than the PDG value \cite{PDG},
as was also obtained in the $N_{f}=2$ Lagrangian in Ref.\ \cite{Paper1}.
However, we obtain values of $m_{K_{1}}$ and $m_{f_{1S}}$ that are very
close to the experimental data, with both mass values deviating from the
PDG data by approximately 20 MeV at the most hence favouring the hypothesis that these states are
predominantly of ${\bar q} q$ nature.

\section{Conclusions}

We have presented a linear sigma model with vector and axial-vector degrees
of freedom that, in our approach, has been extended to $N_{f}=3$.
Implementing the spontaneous symmetry breaking in the model yields not only
the known $\eta _{N}$-$f_{1N}$ and $\vec{\pi}$-$\vec{a}_{1}$ mixings \cite%
{Paper1}\ but also the $\eta _{S}$-$f_{1S}$, $K_{S}$-$K^{\star }$ and $K$-$%
K_{1}$ mixings. Removing the non-diagonal terms in the Lagrangian and
subsequently bringing the $\eta _{N,S}$, $\vec{\pi}$, $K_{S}$ and $K$
derivatives to the canonical form leads us to, among others, define
the kaon renormalisation coefficient $Z_{K}$. Besides its definition
formula, Eq.\ (\ref{ZK1}), $Z_{K}$ also possesses two other formulas
obtained from the difference of the strange axial-vector and vector mass
terms $m_{f_{1S}}^{2}-m_{\omega _{S}}^{2}$ and $m_{K_{1}}^{2}-m_{K^{\star
}}^{2}$, Eqs.\ (\ref{ZK2}) and (\ref{ZK3}). Setting $m_{\rho }$, $%
m_{K^{\star }}$ and $m_{\omega _{S}}$ to their PDG values and enforcing the
equality of the three mentioned $Z_{K}$ formulas yields constraints on $%
m_{a_{1}}$, $m_{K_{1}}$ and $m_{f_{1S}}$. We leave $m_{a_{1}}$ free due to
its large decay width. Consequently, we obtain $1276$ MeV $\leq
m_{K_{1}}\leq 1292$ MeV and $1406$ MeV $\leq m_{f_{1S}} \leq 1423$ MeV which is in
a very good agreement with the experimental data \cite{PDG}. Note that this result
favours the $K_1(1270)$ and $f_1(1420)$ resonances as predominantly ${\bar q} q$ states.\\
Calculations of all meson masses and, subsequently, of
the decay widths of the resonances in the Lagrangian (\ref{Lagrangian}) present an outlook of our approach \cite{PGRKW}.







\bibliographystyle{aipproc}
\bibliography{template-8d}

\IfFileExists{\jobname.bbl}{}  {\typeout{}  \typeout{******************************************}  \typeout{** Please run
"bibtex \jobname" to optain}  \typeout{** the bibliography and then re-run
LaTeX}  \typeout{** twice to fix the references!}  \typeout{******************************************}  \typeout{}  }

\end{document}